\documentclass[runningheads,a4paper]{llncs}

\usepackage[round]{natbib}   

\usepackage[ruled]{algorithm2e}

\SetAlFnt{\small}
\SetAlCapFnt{\small}
\SetAlCapNameFnt{\small}
\SetAlCapHSkip{0pt}
\IncMargin{-\parindent}

\usepackage{amssymb}
\usepackage{graphicx}
\usepackage{subfigure}

\usepackage{bbm}
\usepackage{amsmath}

\usepackage{amsmath}

\usepackage{url}

\usepackage{todonotes}

\begin{document}

\titlerunning{Systemic Risk Management in Financial Networks with CDSs}

%
%

%
\authorrunning{Leduc, Poledna \& Thurner}


%
%

\mainmatter  

\title{Systemic Risk Management in Financial Networks with Credit Default Swaps}


%
%
\title{Systemic Risk Management in Financial Networks with Credit Default Swaps\thanks{Matt V. Leduc can be reached at mattvleduc@gmail.com. Sebastian Poledna can be reached at poledna@iiasa.ac.at. Stefan Thurner (corresponding author) can be reached at stefan.thurner@meduniwien.ac.at.}}

\titlerunning{Systemic Risk Management in Financial Networks with CDSs}

%
%
\author{Matt V. Leduc$^{1}$, Sebastian Poledna$^{1,2,3}$, Stefan Thurner$^{1,2,3,4}$}


\institute{
$^1$IIASA, Schlossplatz 1, 2361 Laxenburg, Austria\\
$^2$Section for Science of Complex Systems, Medical University of Vienna, Spitalgasse 23, 1090 Vienna, Austria\\
$^3$Complexity Science Hub Vienna, Josefst\"adterstrasse 39. 1080 Vienna, Austria\\
$^4$Santa Fe Institute, 1399 Hyde Park Road, Santa Fe, NM 87501, USA
}

\authorrunning{Leduc, Poledna \& Thurner}


%


%
%

\maketitle
 \ \ \  \ \ \  \ \ \  \ \ \  \ \ \  \ \ \  \ \ \  \ \ \  \ \ \  \ \ \  \ \ \  \ \ \  \ \ \  (August 2017)
\\
The published version of this article appeared in the \textit{Journal of Network Theory in Finance}
(2017) and is available at http://dx.doi.org/10.21314/JNTF.2017.034

\begin{abstract}
We study insolvency cascades in an interbank system when banks are allowed to insure their loans with credit default swaps (CDS) sold by other banks. We show that, by properly shifting financial exposures from one institution to another, a CDS market can be designed to rewire the network of interbank exposures in a way that makes it more resilient to insolvency cascades. A regulator can use information about the topology of the interbank network to devise a systemic insurance surcharge that is added to the CDS spread. CDS contracts are thus effectively penalized according to how much they contribute to increasing systemic risk. CDS contracts that decrease systemic risk remain untaxed. We simulate this regulated CDS market using an agent-based model (CRISIS macro-financial model) and we demonstrate that it leads to an interbank system that is more resilient to insolvency cascades.

\end{abstract}

  \
\\
   \
\\
 \textbf{Key Message}
 \begin{itemize}

 \item CDS contracts can rewire the interbank network and make it more resilient to insolvency cascades.
 
 \item CDS contracts can be penalized (e.g. taxed) according to how much they contribute to systemic risk.
 
 \item We simulate such a regulated CDS market and show that it leads to a more resilient interbank system
 
\end{itemize}

\section{Introduction}




Financial derivative contracts have been criticized for their role in the 2007-2008 financial crisis. The opacity of such contracts -- and the fact that they are often unregulated or traded over-the-counter (OTC) -- has drawn criticism, in particular to credit default swaps (CDSs). CDSs were created in the 1990s as a risk management tool by which a loan could be insured against default risk. Their use is however mostly speculative and has drastically increased since the early 2000s. By 2007, the size\footnote{See for example: "ISDA: CDS Marketplace :: Market Statistics". Isdacdsmarketplace.com. December 31, 2010. Accessed on September 4, 2015.} of this market (in terms of outstanding CDS amount) was $\$ 62.2$ trillion and, although it decreased after the financial crisis of 2007-2008, it remained considerable at about $\$ 25.5$ trillion in 2012. 

In a financial system, institutions are interconnected through a complex network of exposures. This complex web of financial exposures creates systemic risk: the insolvency of a particular institution and the resulting default on its loans can precipitate other institutions into insolvency, thereby generating an insolvency cascade. Studying financial systems from a network perspective has thus received a lot of attention in recent years. It is now known that different financial network topologies have different impacts on the probability of systemic collapse (\cite{eisenberg2001systemic,gai2010contagion, boss2004network, battiston2012debtrank,thurner2013debtrank}). In this sense managing systemic risk reduces to the technical problem of re-shaping the topology of financial networks. Properly used, CDSs have the effect of transferring a financial exposure from one institution to another. They can thus be used to change the topology of the financial network of interbank exposures. To understand whether CDSs can be efficiently used for risk management, it is thus essential to study the impact they have on the topology of financial networks. 

In this article, we describe how a regulator can use CDSs to restructure the interbank network. Banks are exposed to each other through interbank loans resulting from the conduct of normal banking operations. To guard against the risk of default of a counter-party on a given loan, a bank can buy a CDS contract from another bank. A regulator can then use information about the topology of the interbank system to impose a systemic insurance `surcharge' that is added to the CDS spread. This surcharge is proportional to the amount of systemic risk created by the contract. This effectively constitutes a mechanism that 'taxes' CDS contracts according to how they contribute to increasing systemic risk. CDS contracts that decrease systemic risk -- by reallocating exposures more efficiently -- remain untaxed. This mechanism has the effect of `matching' CDS counter-parties in a way that reduces systemic risk. With an agent-based model (CRISIS macro-financial model), we demonstrate how this mechanism leads to a self-organized re-structuring of the interbank system that makes it considerably more resilient to insolvency cascades.

One of our contributions is to study the financial system as a multi-layer network. Here the different layers represent different types of contractual obligations (loans, derivatives). We show that insolvency cascades can spread through these different types of edges in non-standard ways. This complements recent treatments of systemic risk in multilayered networks (e.g. \cite{cont2014credit,burkholz2015systemic, poledna2015multi,aymanns2016illiquidity}). Our main contribution is however to propose a mechanism to regulate the CDS market so as to control the formation of the interbank network (modeled as a multi-layer network). The existing literature\footnote{See also \cite{markose2012systemic} for a treatment of systemic risk associated with the CDS market using a network approach and analysis of an eigenvalue centrality-based tax.}, on the other hand, is fairly young and has mainly focused on the introduction of a central clearing house for derivatives (and the associated effects of bilateral netting) or on setting collateral levels (e.g. \cite{duffie2011}). The only other similar mechanisms proposed were in \cite{poledna2014elimination, thurner2013debtrank, leduc2015equilibrium}. We also show that an unregulated CDS market, in which banks are allowed to speculate by buying CDS on loans that they do not own, can drastically increase systemic risk by creating many contagion channels.

The paper is organized as follows. In Section \ref{sec:MultiLayerNet}, we present a multi-layer model of the interbank system, which allows us to study how CDSs affect the topology of exposures between the banks. In Section \ref{sec:RegCDSMarket}, we introduce a measure of systemic risk and we characterize how a particular CDS contract increases or decreases systemic risk. We then discuss how a regulator can design a systemic surcharge mechanism for CDS contracts, which effectively rewires the interbank system. In Section \ref{sec:Sims}, we simulate this mechanism using an agent-based model (CRISIS macro-financial model) and show that it can considerably improve the resilience of the interbank system to  insolvency cascades.

\section{A Multi-Layer Network Model for the Interbank System}
\label{sec:MultiLayerNet}

\subsection{Network of Loan Exposures}
Let us consider a financial system composed of $B$ banks. Based on their liquidity needs, these banks extend loans to each other. Assume that there can be more than one loan between banks $i$ and $j$. Let $\tilde{L}_{ij} = \sum_k l^k_{ij}$, where $l^k_{ij}>0$ is the value of the $k$-th loan extended by $j$ to $i$. We let $l^k_{ij} = - l^k_{ji}$ so that $\tilde{L}_{ij}>0$ means $j$ has a positive net exposure to $i$. This quantity then allows us to define the net exposure matrix as
\begin{equation}
L = \max(0,\tilde{L}).
\end{equation}

\begin{figure*}
  \centerline{
\includegraphics[scale=0.55]{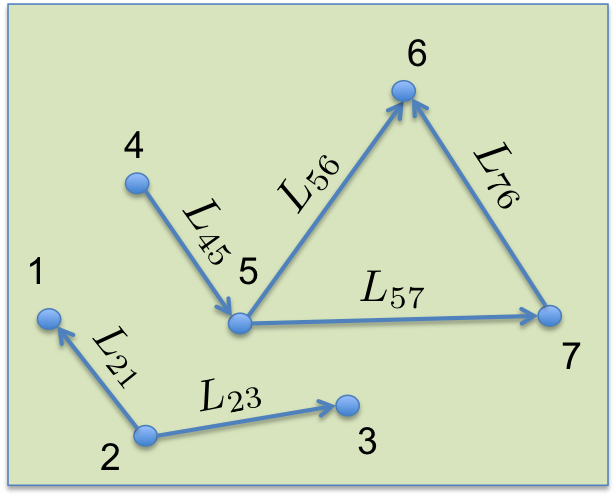}
  }
  \caption{Network of interbank loan exposures. \textit{Here $B=7$ and the interbank system is thus composed of $7$ banks. Each edge represents the net loan exposure between two banks. 
}}
  \label{fig:IB_loans}
\end{figure*}

A positive entry $L_{ij}>0$ represents the net exposure of $j$ to $i$ while $L_{ij}=0$ means that $j$ is not exposed to $i$. The net exposure matrix $L$ defines a \textit{network} of net exposures, as represented in Fig. \ref{fig:IB_loans}.

Loans carry credit risk: in the event of bank $i$'s default, bank $j$ may not be able to recover the net amount $L_{ij}$ that it has lent to bank $i$. The network of loan exposures $L$ creates \textit{systemic risk}. Indeed, the impact of the default of some borrower $i$ not only affects the lender $j$ itself, but potentially the lenders' creditors as well as their own creditors and so on. In the example of Fig. \ref{fig:IB_loans}, the insolvency of bank $4$ not only affects bank $5$, but may also propagate to banks $6$ and $7$, thus generating an insolvency cascade. Likewise, the lender is not only vulnerable to the defaults of his own borrowers, but also to the defaults of his borrowers' borrowers and so on. In an interbank network, credit risk thus ceases to be a local property and becomes \textit{systemic}. Such insolvency cascades have been extensively studied, e.g. \cite{eisenberg2001systemic,gai2010contagion,poledna2014elimination,acemoglu2013systemic,
battiston2012debtrank,
ElliotGolubJackson2014,amini2013resilience}. The size of an insolvency cascade is shown to depend on the topology of the exposures network. Systemic risk must thus be understood as a network property.

In order to guard against the risk of default of a borrower, a lending bank can insure its loan using a financial instrument known as a credit default swap (CDS).

\subsection{Basics of Credit Default Swaps (CDSs)}

\begin{figure}
\begin{minipage}{\columnwidth}
\centering
\includegraphics[width=3.0in]{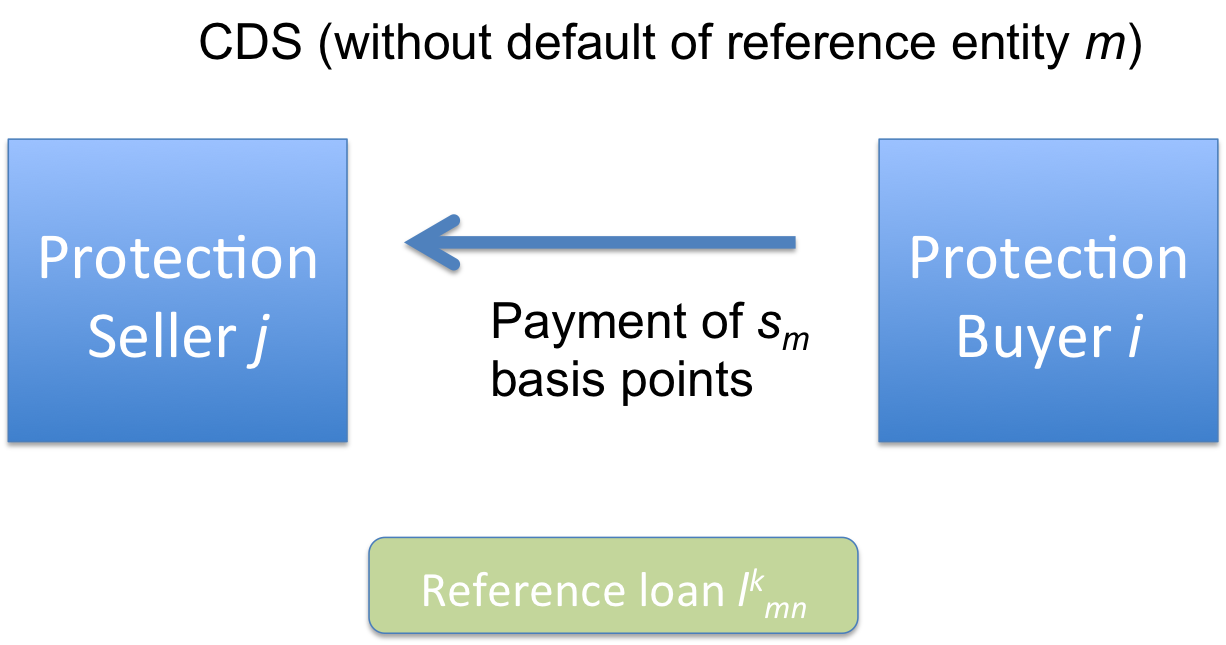}
    \text{(a)}
\end{minipage}
\\
\\
\\
\\
\begin{minipage}{\columnwidth}
\centering
\includegraphics[width=3.0in]{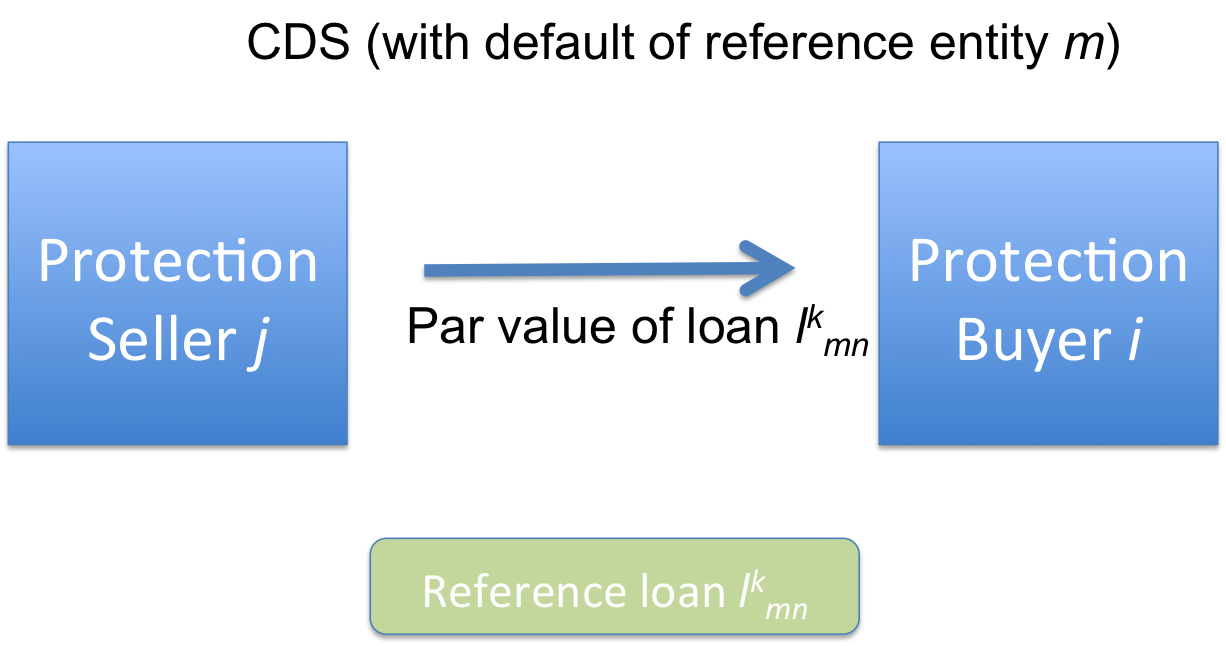}
    \text{(b)}
\end{minipage}
\caption{A typical CDS contract between buyer $i$ and seller $j$ on some reference loan $l^k_{mn}$. (a)  Transactions when there is no default of the reference entity $m$. (b) Transactions in the event of default of the reference entity $m$.}
\label{fig:CDSbasics} 
\end{figure}

A credit default swap (CDS) is a financial swap agreement, in which the buyer makes a series of periodic payments (known as `spread' payments) in exchange for a promise that the seller will compensate the buyer in the event of default on some specific reference loan (or bond). If the CDS buyer holds the reference loan, it is thus a form of insurance contract. In the event of default on that reference loan, the buyer of the CDS typically receives an amount equal to the par value of the loan from the CDS seller. This is illustrated in Fig. \ref{fig:CDSbasics}, where a bank $i$ buys CDS protection from bank $j$ on some reference loan $l^k_{mn}$ extended to bank $m$.

If, in the CDS contract of Fig. \ref{fig:CDSbasics}, the buyer (bank $i$) also owns the reference loan $l^k_{mi}$, then the CDS is used as insurance against the default of $m$. We will refer to such a contract as a `covered' CDS. In reality, the buyer of the CDS need not own the reference loan $l^k_{mn}$. The latter will then be referred to as a `naked' CDS and can allow the parties to speculate on the credit worthiness of the reference entity (bank $m$).



\subsection{Effect of Covered CDSs on the 
Interbank Network Topology}

In this section, we will examine the effect of covered CDSs on the interbank network topology.
 We will assume that the fulfillment of CDS contracts is guaranteed by some well-capitalized regulating agency. The case of naked CDSs will be discussed later.

\subsubsection{CDS Exposures}

Let $\bar{C}_{ij}^{l^k_{mn}}>0$ denote the promised payment of the CDS contract sold by bank $j$ to bank $i$ in the event of default of the reference entity $m$ on the reference loan $l^k_{mn}$. We can now define $\tilde{C}^{m}_{ij} = \sum_{k,n} \bar{C}_{ij}^{l^k_{mn}}$ as the total promised payments on CDS contracts sold by bank $j$ to bank $i$ in the event of default of $m$. We let $\tilde{C}^{m}_{ij} = - \tilde{C}^{m}_{ji}$ so that $\tilde{C}^{m}_{ij} > 0$ means $j$ has a positive net CDS exposure to $i$. This quantity then allows us to define the net CDS exposure matrix on the reference entity $m$:

\begin{equation}
C^{m} = \max(0,\tilde{C}^{m}).
\label{eq:CDS_m}
\end{equation}

A positive entry $C^m_{ij}>0$ represents the net CDS exposure of $j$ to $i$ on the reference entity $m$. In the event of $m$'s default, $j$ will thus have to pay this amount to $i$. $C^m_{ij}=0$ means that $j$ has no CDS exposure to $i$ on the reference entity $m$.




\subsubsection{A Two-Layer Representation for the Interbank Network}
Note that the collection of all CDS exposure matrices $\{C^m\}_{m =1,...,B}$ defines a multiplex network, i.e. a network where different types of edges may exist between different banks. Each particular type of edge represents a net CDS exposure on a particular reference entity $m$. The edge type is thus labeled by $m$.

\begin{figure*}
  \centerline{
\includegraphics[scale=0.3]{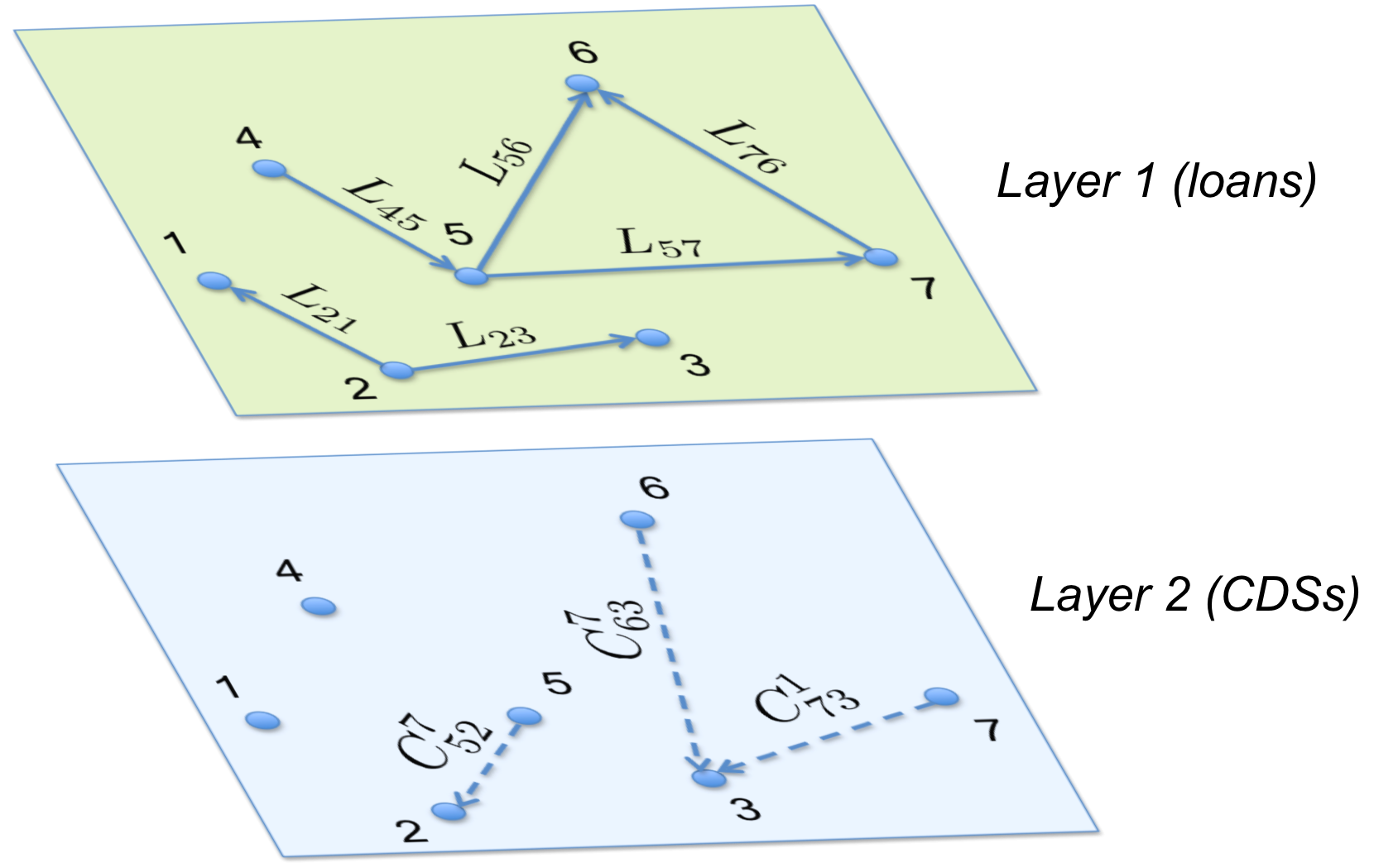}
  }
  \caption{A two-layer network representing the interbank system. \textit{Here $B=7$ and the interbank system is thus composed of $7$ banks. Layer $1$ represents net loan exposures between those banks. Layer $2$ represents net CDS exposures on various reference entities (banks).}}
  \label{fig:MultiLayer}
\end{figure*}

The interbank system can thus be represented as a two-layer network: The first layer represents the network of loan exposures $L$ between the $B$ banks, while the second layer represents the multiplex network of CDS exposures between the $B$ banks. Such a two-layer representation is similar to that used in \cite{poledna2015multi}. It is represented in Fig. \ref{fig:MultiLayer}. 
From that figure, we see that a directed edge $ij$ of weight $L_{ij}$ on layer $1$ represents the net loan exposure of bank $j$ to bank $i$. On the other hand, a directed edge $ij$ of weight $C^m_{ij}$ represents the net amount that bank $j$ will have to pay bank $i$ in the event of the default of bank $m$. 

Note that the relations between the different layers are highly non-standard. The default of a particular bank will not only affect its creditors, but also the banks that have sold CDS contracts on that reference bank. It may therefore be difficult to measure the impact of the bankruptcy of one bank on the system as it depends on the interaction between the layers. If we assume that the fulfillment of CDS contracts is guaranteed by some well-capitalized regulating agency, then in the case of covered CDSs this multi-layer network can be conveniently mapped to a single-layer network, representing the \textit{effective net exposures}\footnote{Note that we allow here the netting of different asset classes. As a matter of fact there are detailed international agreements on the netting procedure in the case of failure of a counter-party. Typically, for deposits and loans it is customary to calculate the gross exposure instead of the net exposure and derivatives can only be netted by each type of derivative contract.} of banks to one another. This effective net exposure is defined, for any banks  $i$ and $j$, as
\begin{equation}
L^{\rm{eff}}_{ij} = \max \Big(0,L _{ij} - \sum_z C^i_{jz} + \sum_{z^{'}} C^i_{z^{'}j} \Big).
\label{eq:L_eff}
\end{equation}

\begin{figure*}
  \centerline{
\includegraphics[scale=0.45]{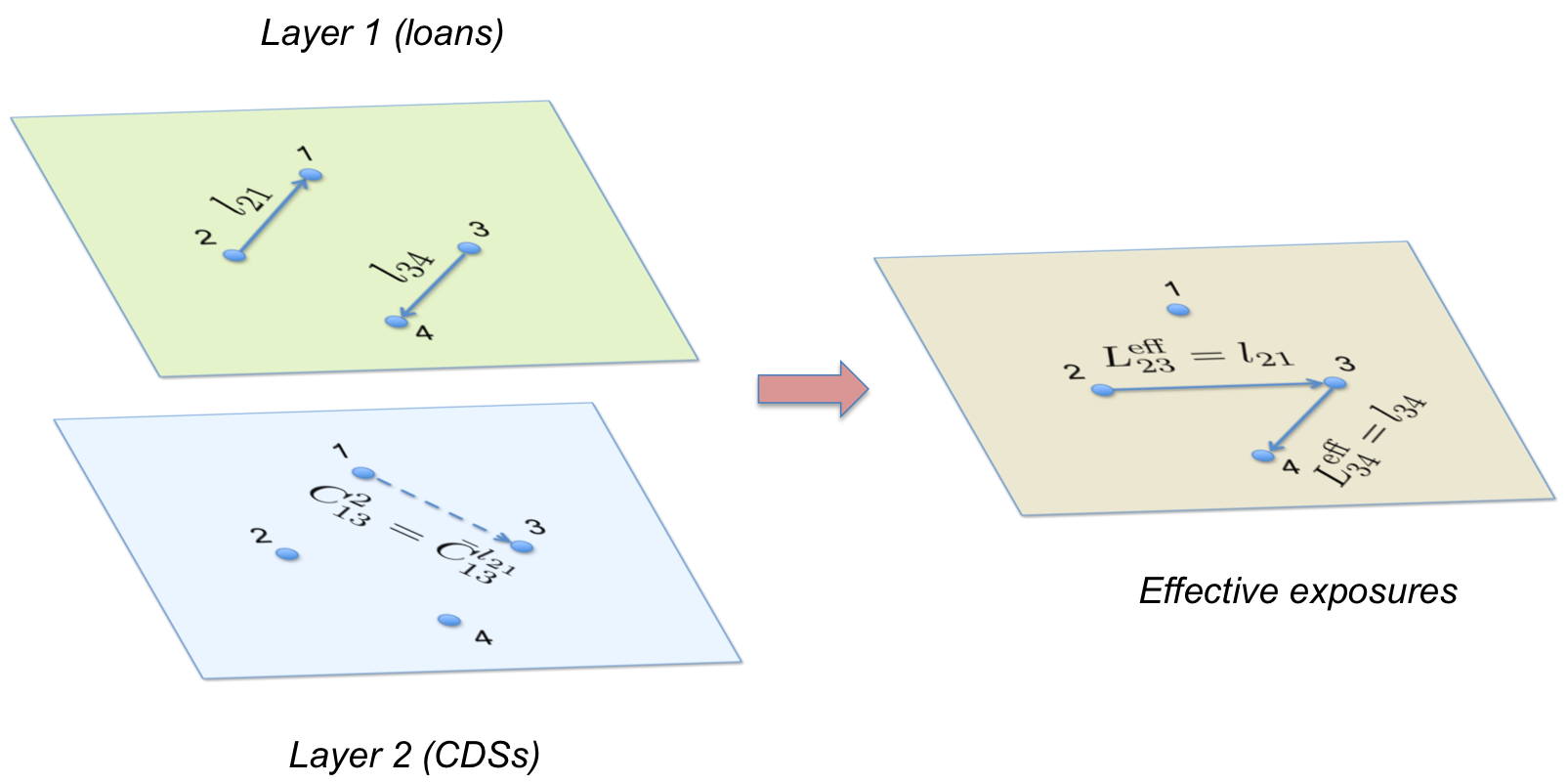}
  }
  \caption{Effective Exposure with a `Covered' CDS: the Topology of the Loans Exposure Network is 'Rewired'. \textit{We assume that the fulfillment of CDS contracts is guaranteed by a well-capitalized regulating agency. In this simple example, there are $4$ banks $(B=4)$ and only one CDS exposure on reference entity $2$. Since the buyer of the CDS $\bar{C}^{l_{21}}_{13}$ (bank $1$) also owns the loan $l_{21}$, the CDS contract transfers this loan exposure to the seller of the CDS (bank $3$).}}
  \label{fig:eff_exp_CDS}
\end{figure*}

This mapping of the two layers into a single exposure layer is illustrated in Fig. \ref{fig:eff_exp_CDS}.

In Fig. \ref{fig:eff_exp_CDS}, we see that covered CDS contracts have the effect of rewiring the network of loan exposures by transferring exposures from one bank to another. In this simple example, there are $4$ banks and only one CDS exposure on reference entity (bank) $2$. Thus bank $1$ has bought a CDS from bank $3$ on the reference bank $2$. The value of the CDS exposure here is $l_{21}$ so that bank $1$ has insured its net loan exposure to bank $2$ against the possible default of bank $2$. This loan exposure $l_{21}$ is thus effectively transferred to $3$. It is important to emphasize that in Fig. \ref{fig:eff_exp_CDS}, the buyer of the CDS (bank $1$) also owns the reference loan $l_{21}$. 

\subsection{Effect of Naked CDSs on the Interbank Network Topology}
\label{sec:effectNakedCDS}
\begin{figure*}
  \centerline{
\includegraphics[scale=0.45]{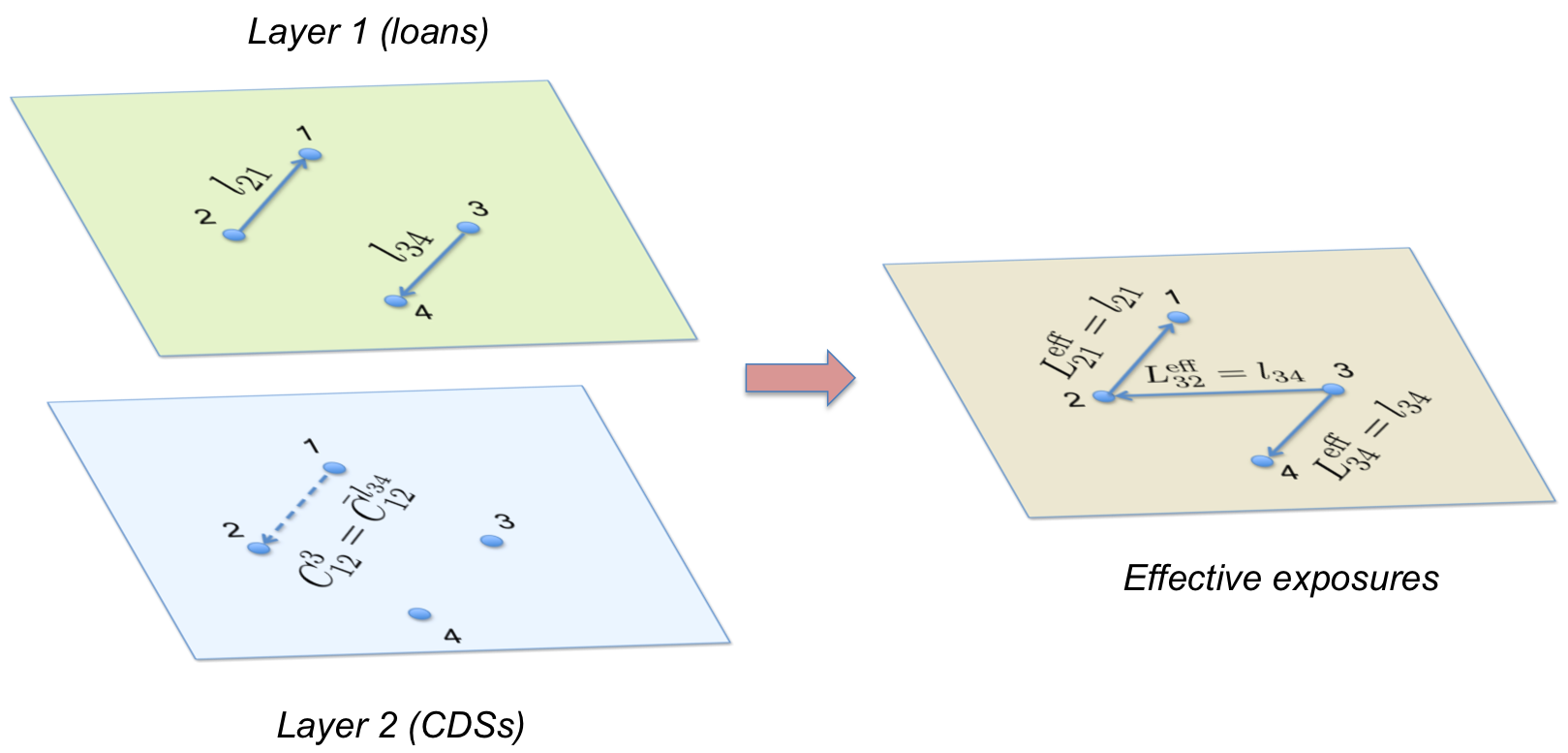}
  }
  \caption{Effective Exposure with a `Naked' CDS: an Additional Exposure (Edge) is Created. \textit{We assume that the fulfillment of CDS contracts is guaranteed by a well-capitalized regulating agency. In this simple example, there are $4$ banks $(B=4)$ and only one CDS exposure on reference entity $3$. Since the buyer of the CDS contract $\bar{C}^{l_{34}}_{12}$ (bank $1$) does not own the loan $l_{34}$, the CDS contract creates a new exposure:  the seller of the CDS (bank $2$) is now exposed to the reference entity (bank $3$). The effective exposure layer however does not take into account the positive payment to bank $1$ in the event of the failure of bank $3$. Bank $1$'s equity must be adjusted `off network'.}
  }  
  \label{fig:eff_exp_naked_CDS}
\end{figure*}

As we saw previously, a \textit{covered} CDS is a contract $\bar{C}^{l^k_{mi}}_{ij}$ in which the buyer of protection (i.e. bank $i$) also owns the underlying loan $l^k_{mi}$. This CDS contract is thus a form of insurance that $i$ buys from $j$ against the default of $m$ on loan $l_{mi}^k$. On the other hand, a \textit{naked} CDS contract $\bar{C}^{l^k_{mn}}_{ij}$, is a contract in which the buyer (i.e. bank $i$) does not own the underlying loan $l^k_{mn}$. The buyer therefore does not buy insurance protection, but rather speculates on the credit worthiness of bank $m$ and hopes to profit in the event of $m$'s default. The effect of such contracts on the interbank network topology is more complex\footnote{The difficulty of studying naked CDSs in interbank systems has been noted in \cite{schuldenzucker2016clearing}. The authors study the problem of clearing a system of interconnected banks when banks can also enter into CDSs contracts. They show how difficult it may be to find a clearing payment vector, which may not even exist. They also show that banning naked CDSs can guarantee existence of a unique clearing vector.} and the transformation of the two-layer representation into an effective exposure network is not sufficient to capture it. In Fig. \ref{fig:eff_exp_naked_CDS}, we show the effect of a naked CDS on the network of effective exposures. In this case, bank $1$ buys a CDS on a loan that it does not own (i.e. $l_{34}$). The effect is the creation of a new edge in the network of effective exposures. Thus instead of rewiring the network, a naked CDS has the potential to create new contagion channels. Note however that in the event of a default of the underlying entity (bank $3$), bank $1$ will receive a positive payment, which will positively affect its equity. This adjustment to bank $1$'s equity has to be performed 'off network', as it is not captured by the effective exposure network.

\section{A Regulated CDS Market}
\label{sec:RegCDSMarket}
\subsection{Effect of Covered CDSs on Systemic Risk}

As a network property, systemic risk can be quantified by network metrics. One such metric is DebtRank (\cite{battiston2012debtrank, thurner2013debtrank, poledna2014elimination}), which is a recursive method to determine the systemic relevance of nodes within a financial network. DebtRank has a natural interpretation as the fraction of the total economic value in the financial network that is potentially lost as a result of the default (or distress) of a bank (or a set of banks). Considering the network of effective exposures $L^{\rm{eff}}$ and the capital $E = [E_1, ..., E_B]$ of all banks, the DebtRank $R_i(L^{\rm{eff}},E)$ of bank $i$ can be readily computed (see \cite{battiston2012debtrank} for details).

We study an environment in which only covered CDS contracts are allowed and are guaranteed by some well-capitalized regulating agency. If bank $i$ defaults and cannot honor its obligations, then bank $j$ may lose the value $L_{ij}$ of its net loan exposure to bank $i$. However if some bank $z$ has a CDS exposure $C^{i}_{jz}$ to bank $j$ on the reference bank $i$, then it will pay bank $j$ an amount to compensate the loss. If bank $z$ is unable to fully pay this amount because it exceeds its capital, then bank $z$ will go bankrupt. The regulating agency will however step in to guarantee the CDS payment to bank $j$. The effective net exposure $L^{\rm{eff}}$ is thus the appropriate quantity to measure the exposure of one institution to another. The assumption that there is a well-capitalized regulating agency guaranteeing CDS contracts has already been made in the literature. For example, see \cite{duffie2011, cont2014credit} for papers that study the central clearing of derivative contracts.

We can now naturally define the \textit{expected systemic loss} as
\begin{equation}
EL^{\rm{syst}}(L^{\rm{eff}},E) = \sum_{h=1}^{B}P_h^{\rm{def}} V R_h(L^{\rm{eff}},E)
\end{equation}
where $P_h^{\rm{def}}$ is the exogenous probability of default of bank $h$ and $V$ is the combined economic value of all banks. $R_h(L^{\rm{eff}},E)$ is the DebtRank of bank $h$, as introduced earlier. By \em exogenous \em probability of default, we mean that $P_h^{\rm{def}}$ is the probability that bank $h$ fails due to factors other than the default of other banks to which it is exposed. This allows us to separate exogenous (or non-network) effects from contagion (or network) effects.

To calculate the contribution of a particular CDS contract $\bar{C}^{l^k_{mi}}_{ij}$ to the expected systemic loss, we need to adjust layer $m$ by updating the matrix $\tilde{C}^{m}$ as follows 

\begin{equation}
\tilde{C}^{m'}_{ij} = \tilde{C}^{m}_{ij} + \bar{C}^{l^k_{mi}}_{ij}
\end{equation}
and
\begin{equation}
\tilde{C}^{m'}_{ji} = \tilde{C}^{m}_{ji} - \bar{C}^{l^k_{mi}}_{ij}
\end{equation}
and then recalculating $L^{\rm{eff}}$ by using equations (\ref{eq:CDS_m})-(\ref{eq:L_eff}). We denote the effective net exposure matrix in the presence of that specific CDS contract by $L^{\rm{eff}[+\bar{C}^{l^k_{mi}}_{ij}]}$.
It is now straightforward to compute the marginal effect of that particular CDS contract on the expected systemic loss:
\\
\begin{eqnarray}
\Delta^{\rm [+ \bar{C}^{l^k_{mi}}_{ij}]} EL^{\rm{syst}} &=& EL^{\rm{syst}}(L^{\rm{eff}[+ \bar{C}^{l^k_{mi}}_{ij}]},E^{\rm [+\bar{C}^{l^k_{mi}}_{ij}]}) - EL^{\rm{syst}}(L^{\rm{eff}},E) \\
&=& \sum_{h=1}^B P_h^{\rm{def}}  \Big( V^{\rm  [+\bar{C}^{l^k_{mi}}_{ij}]} \cdot R_h(L^{\rm{eff} [+ \bar{C}^{l^k_{mi}}_{ij}]} ,E^{\rm [+\bar{C}^{l^k_{mi}}_{ij}]}) - V \cdot R_h(L^{\rm{eff}},E )    \Big) \nonumber
\end{eqnarray}
where $E^{\rm [+\bar{C}^{l^k_{mi}}_{ij}]}$ is the vector of the banks' equities in the presence of the CDS contract $\bar{C}^{l^k_{mi}}_{ij}$. Likewise, $V^{\rm  [+\bar{C}^{l^k_{mi}}_{ij}]}$ is the combined economic value of all banks in the presence of the CDS contract $\bar{C}^{l^k_{mi}}_{ij}$.

Note that $\Delta^{\rm [+ \bar{C}^{l^k_{mi}}_{ij}]} EL^{\rm{syst}}$ may be either positive or negative. Indeed, a CDS contract may \em decrease \em systemic risk if it shifts a loan exposure from a  bank with higher DebtRank to a  bank with lower DebtRank. The default of the debtor bank then has a smaller impact on the system. In such a case, $\Delta^{\rm [+ \bar{C}^{l^k_{mi}}_{ij}]} EL^{\rm{syst}}$ will be negative.

\subsection{Using Covered CDSs to Rewire the Interbank Network}
\label{sec:UsingCDStoRewire}

The spread that a buyer $i$ of a CDS contract must pay to a seller $j$ on the reference loan $l^k_{mi}$ is a function of the reference entity $m$'s probability of default. We denote it by $s_m$ to emphasize the dependence on $m$. Since a CDS contract promises a payment equal to the face value of the loan, the total spread payment will thus be $s_m \cdot l_{mi}^k$.

While the case for a well-capitalized regulating agency to guarantee the fulfillment of derivative contracts has already been studied in the context of central clearing (e.g. \cite{duffie2011}, \cite{cont2014credit}), here we employ a different approach: a well-capitalized agency regulates the CDS market by forbidding naked CDSs and by incentivizing the covered CDS contracts that decrease systemic risk and penalizing those that increase it. It can do so by adding a \textit{systemic insurance surcharge} to the spread that is normally paid on such a contract. Since a regulator with knowledge of the network topology can compute the marginal effect of a CDS contract on systemic risk, she can devise this systemic surcharge as follows

\begin{equation}
\label{eq:surcharge}
\tau_{ij}(l_{mi}^k) = \zeta \cdot \max \Big[ 0, \int_0^T  v(t) \cdot \Delta^{\rm [+ \bar{C}^{l^k_{mi}}_{ij}]} EL^{\rm{syst}}(t) \ dt \Big]
\end{equation}
where 

\begin{equation*}
\Delta^{\rm [+ \bar{C}^{l^k_{mi}}_{ij}]} EL^{\rm{syst}}(t) = \sum_h \hat{p}_h(t) \Big( V^{\rm  [+\bar{C}^{l^k_{mi}}_{ij}]} \cdot R_h(L^{\rm{eff} [+\bar{C}^{l_{mi}^k}_{ij}]} ,E^{\rm [+\bar{C}^{l_{mi}^k}_{ij}]}) 
 - V \cdot R_h(L^{\rm{eff} } ,E)  \Big).
\end{equation*}
 
Here $\zeta > 0$ is some chosen scaling parameter, $v(t)$ is a discount factor and $\hat{p}_h(t)$ is the density function for the exogenous default probability. $T$ is the CDS contract's maturity.

The spread that bank $i$ must pay for a CDS contract sold by  bank $j$ on a loan $l_{mi}^k$ now depends not only on the reference entity $m$, but also on the two parties $i$ and $j$ in the CDS contract, i.e. 
\begin{equation}
s_{ij}(l_{mi}^k) = s_m + \tau_{ij}(l_{mi}^k).
\end{equation}
The effect of that systemic surcharge will be that a buyer $i$ of CDS protection will now  choose the seller $j$ with the smallest effective spread $s_{ij}(l_{mi}^k)$ and thus the contract that contributes the least to increasing systemic risk. Without that systemic surcharge, the buyer would pay the same spread $s_m$ to any seller and thus would be indifferent as to which seller it buys from. Indeed, the spread of a CDS depends solely on the reference entity's probability of default, not on the CDS seller. While the normal spread $s_m$ is collected by the CDS seller, the systemic surcharge $\tau_{ij}(l^k_{mi})$ is collected by the regulating agency and placed in a fund that will be used to guarantee the fulfillment of CDS contracts if the seller were to become insolvent.

\section{Simulation Results and Discussion}
\label{sec:Sims}

\subsection{The Agent-Based Model}
We simulate this regulated CDS market using an agent-based model (CRISIS macro-financial model). This is an economic simulator that combines a well-studied macroeconomic ABM (\cite{poledna2014elimination,delli2008emergent,gatti2011macroeconomics,gaffeo2008adaptive}) with an ABM of the interbank system. We use a modified version of the ABM in \cite{gatti2011macroeconomics}, which we augmented with a multi-layer interbank market. We also follow the same simulation procedure as in \cite{poledna2014elimination}, but with the addition of the CDS market. In \cite{poledna2014elimination}, only a loans market was considered. This is a closed economic system, i.e. it does not allow for cash in-flows or out-flows. For a full description of this agent-based model, the reader is referred to \cite{gatti2011macroeconomics,gualdi2015tipping}.

\begin{figure}
\centering

\includegraphics[width=.6\textwidth]{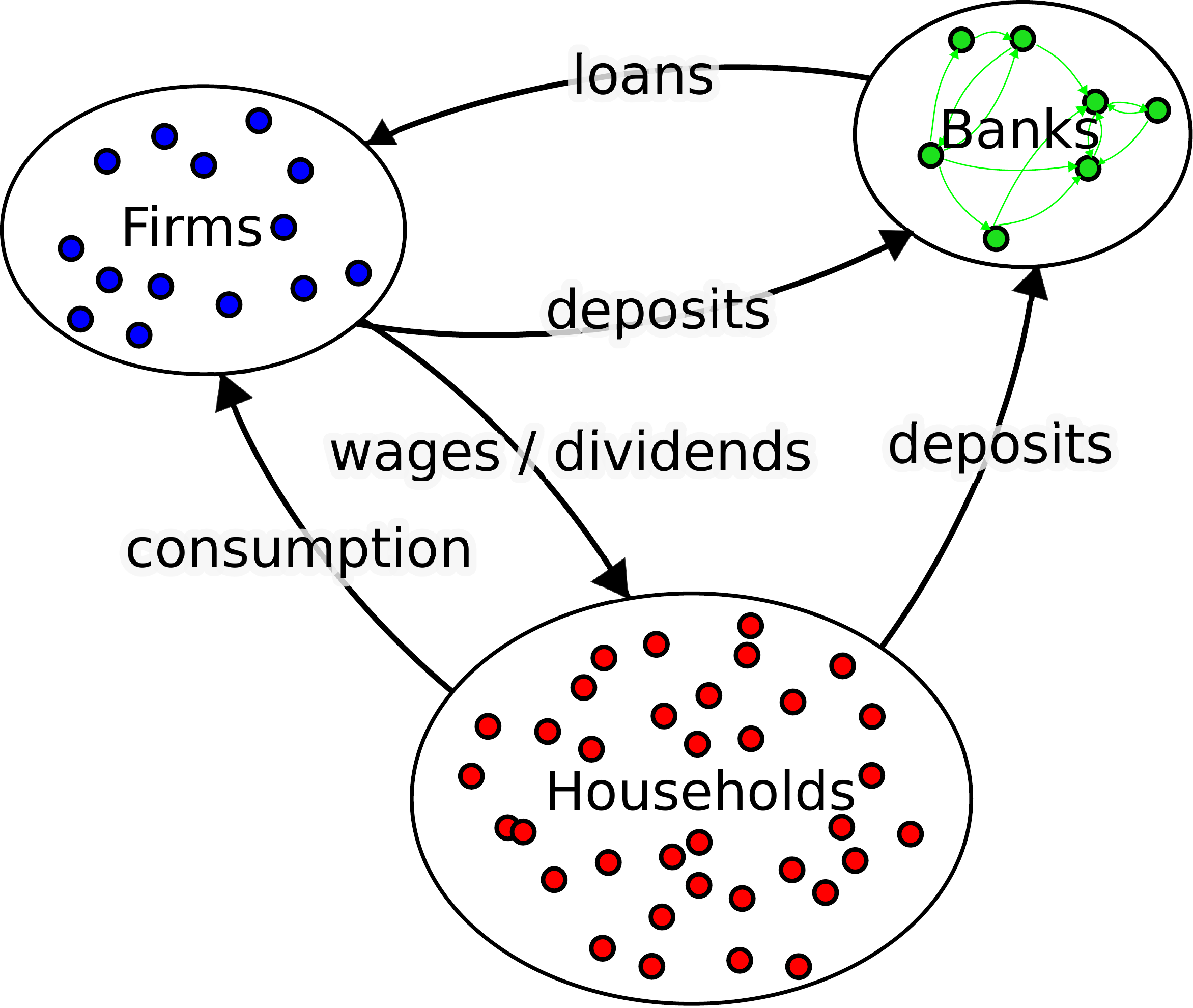}

\caption{Schematic Overview of the Structure of the Agent-Based Model (ABM). \textit{The three agent types (banks, firms, and households) and their interactions are shown. Firms pay dividends to their owners, and wages (financed through income and loans) to their workers. Households consume goods produced by the firms. Households and firms deposit money in banks, banks grant loans to the firms. Most importantly, banks extend inter-bank loans to each other and they can insure them with CDSs issued by other banks.}}

\label{ABM_model}
\end{figure}

The model features three types of agents: households, banks, and firms, as depicted in Fig. \ref{ABM_model}. They interact on five markets

(i) Firms and banks interact on the \textit{credit market}. 

(ii) Banks interact with banks on the \textit{interbank market for loans}. 

(iii) Banks interact with banks on the \textit{interbank market for CDSs}.

(iv) Households and firms interact on the \textit{job market}.

(v) Households and firms interact on the \textit{consumption
goods market}.

\subsubsection{Households}
There are $H$ households, which come in two types: firm owners and workers. Each of them has a personal account $A_{j,b}(t)$ at one of the $B$ banks. $j$ indexes the worker, $b$ the bank. Household accounts are randomly assigned to banks. Workers apply for jobs at the $F$ different firms. If hired, they receive a fixed income $w$ per time step, and supply a fixed labor productivity $\alpha$. Firm owners receive their income through dividends from their firm's profits. At every time step, each household spends a fixed percentage $c$ of its current account on the goods market. They compare prices of goods from $z$ randomly chosen firms and buy the cheapest.

\subsubsection{Firms}
There are $F$ firms producing perfectly substitutable goods. At every time step, firms compute an expected demand $d_i(t)$, and an estimated price $p_i(t)$ (subscript labels the firm), based on a rule that takes into account both excess demand/supply and the deviation of the price $p_i(t - 1)$ from the average price in the previous time step (\cite{gatti2011macroeconomics}). Each firm computes the number of required workers to supply the expected demand. If the wages for the respective workforce exceed the firm's current liquidity, it applies for credit. Firms approach $n$ randomly chosen banks and choose the credit with the most favorable rate. If this rate exceeds a threshold rate $r_{max}$, the firm only asks for $\phi$ percent of the originally desired loan volume. Based on the outcome of this credit request, firms re-evaluate the needed workforce, and hire or fire the needed number of workers. Firms sell the goods on the consumption goods market. Firms go bankrupt if they have negative liquidity after the goods market closes. Each of the bankrupted firm's creditors (banks) incurs a capital loss in proportion to their investment in the company. Firm owners of bankrupted firms are personally liable, their account is divided among the creditors pro rata. They immediately start a new company, with initially zero equity. Their initial estimates for $d_i(t)$ and $p_i(t)$ equal the respective current averages in the population.

\subsubsection{Banks}
There are $B$ banks that offer firm loans at rates that take into account the individual specificity of banks (modeled by a uniformly distributed random variable), and the firms' creditworthiness. Firms pay a credit risk premium according to their creditworthiness that is modeled by a monotonically increasing function of their financial fragility (\cite{gatti2011macroeconomics}). Banks try to provide requested firm loans and grant them if they have enough liquid resources. If they do not have enough cash, they approach other banks in the interbank market to obtain the needed amount. If a bank does not have enough cash and cannot raise the full amount for the requested firm loan on the IB market it does not pay out the loan. Interbank and firms loans have the same duration. Additional refinancing costs of banks remain with the firms. 

At each time step, firms and banks re-pay a fixed percentage of their out-standing debt (principal plus interest). If banks have excess-liquidity they offer it on the interbank market for a nominal interest rate. The interbank relation network is modeled as a weighted, fully-connected network and banks choose the interbank offers with the most favorable rate. Interbank rates $r_{im}$ offered by bank $i$ to bank $m$ take into account the specificity of bank $i$, and the creditworthiness of bank $m$.

Every time a new loan has been extended from one bank to another, the lending bank can insure its loan on the interbank market for CDSs. It can do so by asking another bank (other than the one it just lent money to) to act as counter-party in a CDS contract. Any other bank will demand the same spread $s_m$ for offering protection on an underlying loan extended to bank $m$. This spread reflects the creditworthiness of bank $m$. In an unregulated CDS market, a bank $i$ will thus be indifferent as to which other bank $j$ it buys a CDS from. In a regulated CDS market, on the other hand, a bank will try to buy a CDS from the bank $j$ associated with the lowest effective spread $s_{ij}(l^k_{mi})$ (cf. Section \ref{sec:UsingCDStoRewire}).

If a bank goes bankrupt, the respective creditor bank writes off the respective outstanding loans as defaulted credits. It then receives the promised compensation payment from the bank that issued CDS protection on the defaulted bank. If the issuing bank has not enough equity capital to cover these CDS payments, it defaults. The CDS payments are however guaranteed by the regulating agency, which will step in to cover whatever portion of the CDS payment the issuing bank is not able to make.

Following a bank default, an iterative default-event unfolds for all interbank creditors and all issuers of CDSs. This may trigger a cascade of bank defaults. For simplicity, we assume no recovery for interbank loans. This assumption is reasonable in practice for short-term liquidity. A cascade of bankruptcies happens within one time step. After the last bankruptcy is taken care of the simulation is stopped.

\subsection{Implementation of the Systemic Surcharge}
In the regulated CDS market that we study, only covered CDSs are allowed. The systemic surcharge that is added to all these CDS transactions is implemented as follows. The surcharge is calculated according to Eq. (\ref{eq:surcharge}). Before entering a CDS transaction, a bank $i$ can get quotes of the effective spreads for various counter-parties from the regulating agency. Bank $i$ chooses to buy a CDS  from the bank $j$ with the smallest effective spread $s_{ij}(l^k_{mi})$. Banks thus have an incentive to enter into CDS transactions that either have the effect of reducing systemic risk or have the smallest possible effect on increasing it. In the latter case, the regulating agency collects the surcharge $\tau_{ij}(l^k_{mi})$, which it stores in a fund. This fund may then be used to guarantee the fulfillment of CDS contracts in the event of the bankruptcy of the CDS seller.  

\subsection{Results}
We implement the agent-based model described above for $B=20$ banks, $F=100$ firms and $H=1300$ households. 

For purposes of comparison, we study four different scenarios: (i) An interbank loans market \textit{without} CDSs; (ii) an interbank loans market \textit{without} CDSs but regulated with a Tobin tax on all loan transactions; (iii) an interbank loans market \textit{with unregulated, naked} CDSs;  (iv) an interbank loans market \textit{with covered} CDSs \textit{regulated} as previously described.

Results are averaged over $10,000$ independent, identical simulation runs across $500$ time steps. We set $P_i^{\rm{def}}=0.01$, and $\zeta=0.02$. In the case of a Tobin tax we impose a constant tax rate of $0.2\%$ of the transaction on all interbank rates $r_{im}$ on offer.

\begin{figure}
\centering

\includegraphics[width=.9\textwidth]{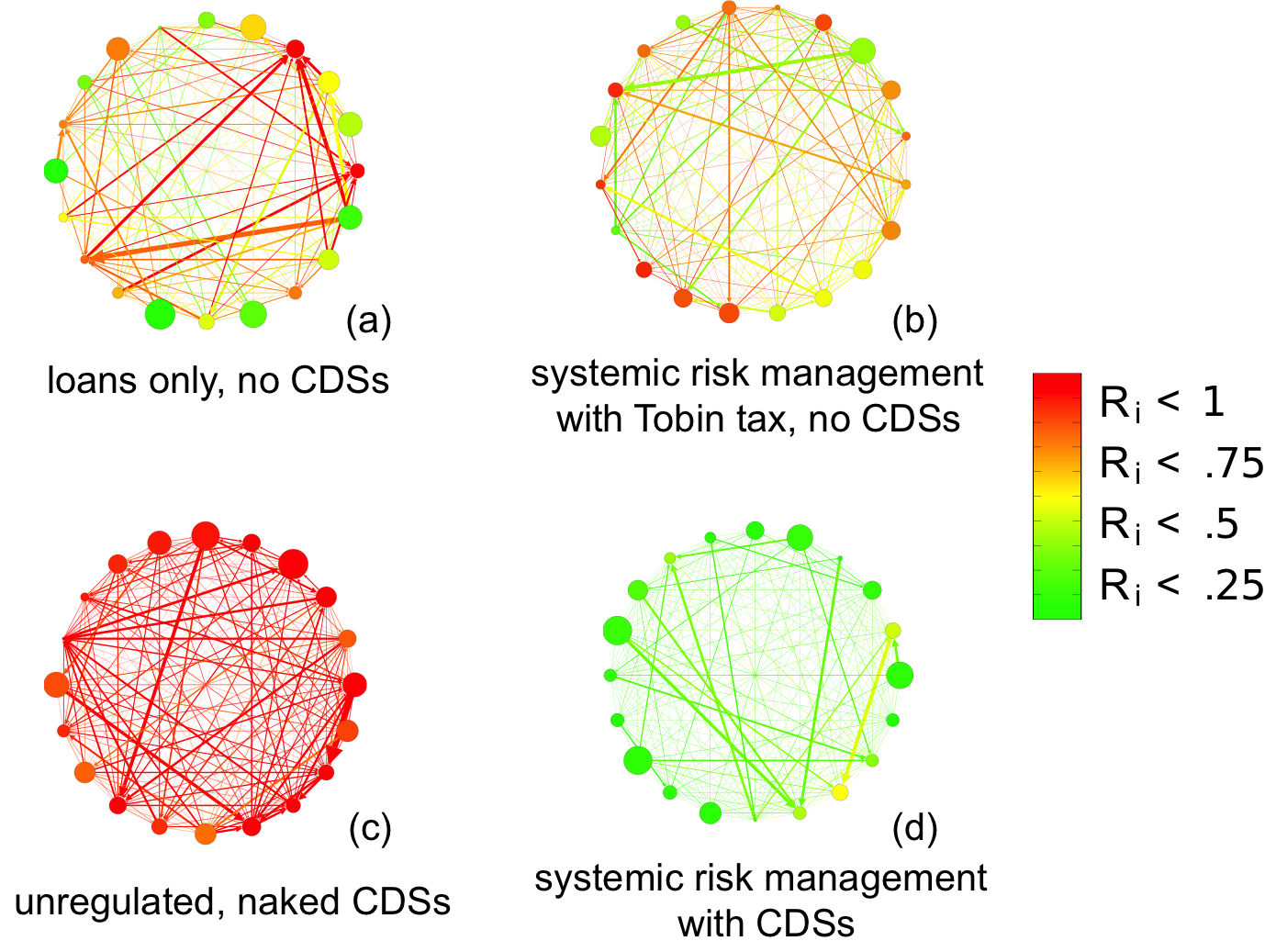}

\caption{Network of Effective Exposures under Different Scenarios.  \textit{The case of an interbank system composed of $20$ banks without a CDS market (loans only) is shown in (a). The case of this interbank system without a CDS market (loans only), but with loans penalized by a Tobin tax is shown in (b). The case of this interbank system with an unregulated (see Footnote \ref{GGG}), naked CDS market is shown in (c), while the case with a regulated, covered CDS market, where systemic risk is managed using the systemic surcharge is shown in (d). The systemic importance $R_i$ of a bank is measured by its DebtRank and color coded. The contribution of an exposure (edge) to systemic risk is also color coded.}}

\label{circular_nets}
\end{figure}

In Fig. \ref{circular_nets}, we plot the networks representing the effective exposures between banks in the four scenarios described above. In Fig. \ref{circular_nets} (a), we see the interbank network without a CDS market. The exposures are thus just loan exposures. We notice that many banks have high systemic importance, due to the absence of incentives to form a network with low systemic risk. Systemic importance is measured by a bank's DebtRank $R_i$, which is color coded. In Fig. \ref{circular_nets} (b), we see that a Tobin transaction tax fails to decrease the systemic importance of banks. Indeed, loans are all taxed equally, which has no effect on controlling the network structure.
In Fig. \ref{circular_nets} (d), on the other hand we see that a regulated, covered CDS market, where systemic risk is managed according to the surcharge mechanism described earlier, allows for a self-organized restructuring of the interbank system in which each bank has considerably lesser systemic importance. In Fig. \ref{circular_nets} (c), however, we see that an unregulated CDS market allowing for speculation (naked CDSs) considerably increases the systemic importance\footnote{\label{GGG} Note that we compute DebtRank using the effective exposure matrix. In the case of naked CDSs, this does not consider positive CDS payments made in the event of the default of the reference entity, as explained in Section \ref{sec:effectNakedCDS}. DebtRank is thus only an approximation in the case of naked CDSs.} of the banks and also the density of the network (i.e. it creates more contagion channels).

\begin{figure}
\centering
\includegraphics[width=.7\textwidth]{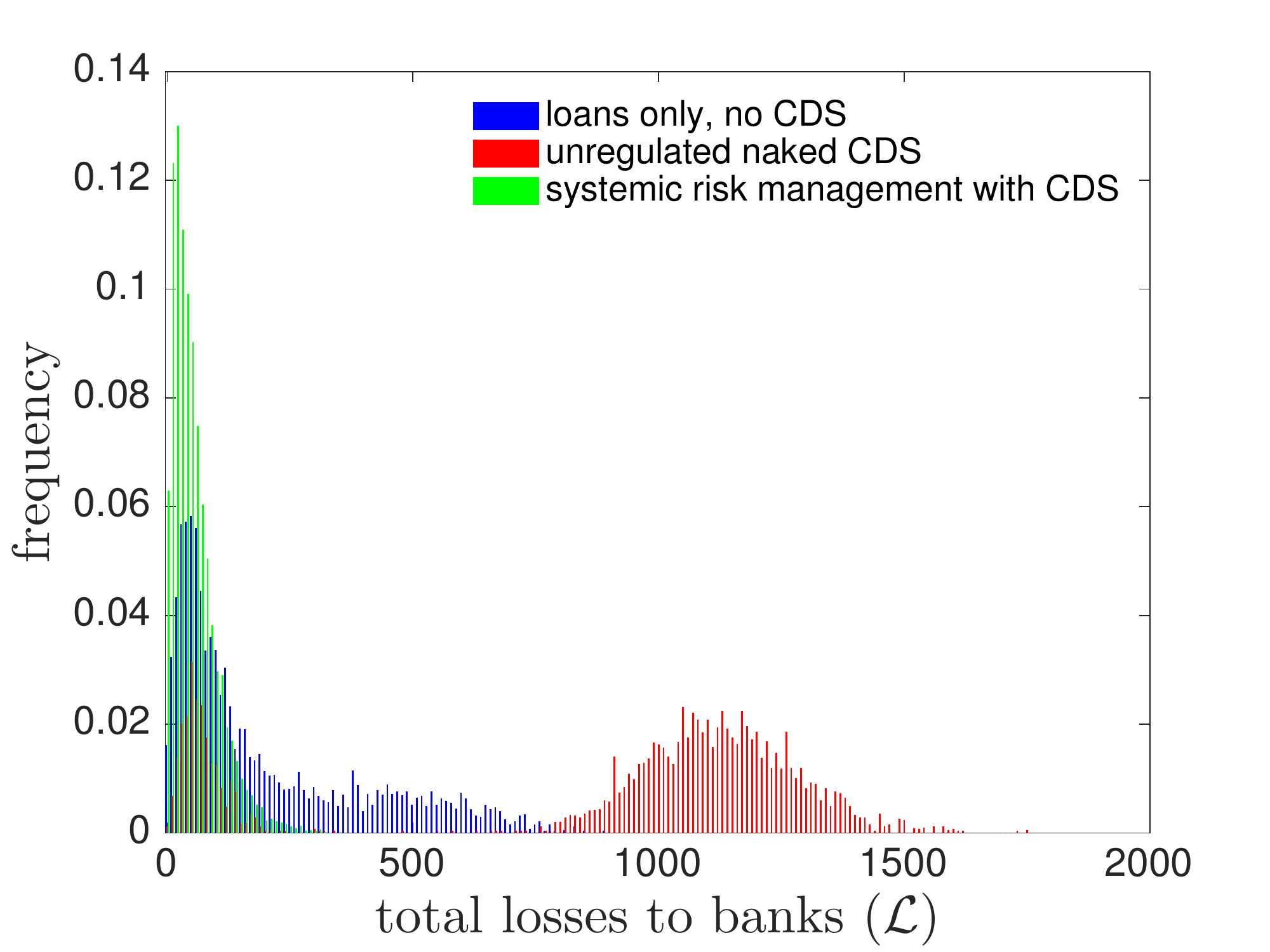}
\text{(a)}
\includegraphics[width=.7\textwidth]{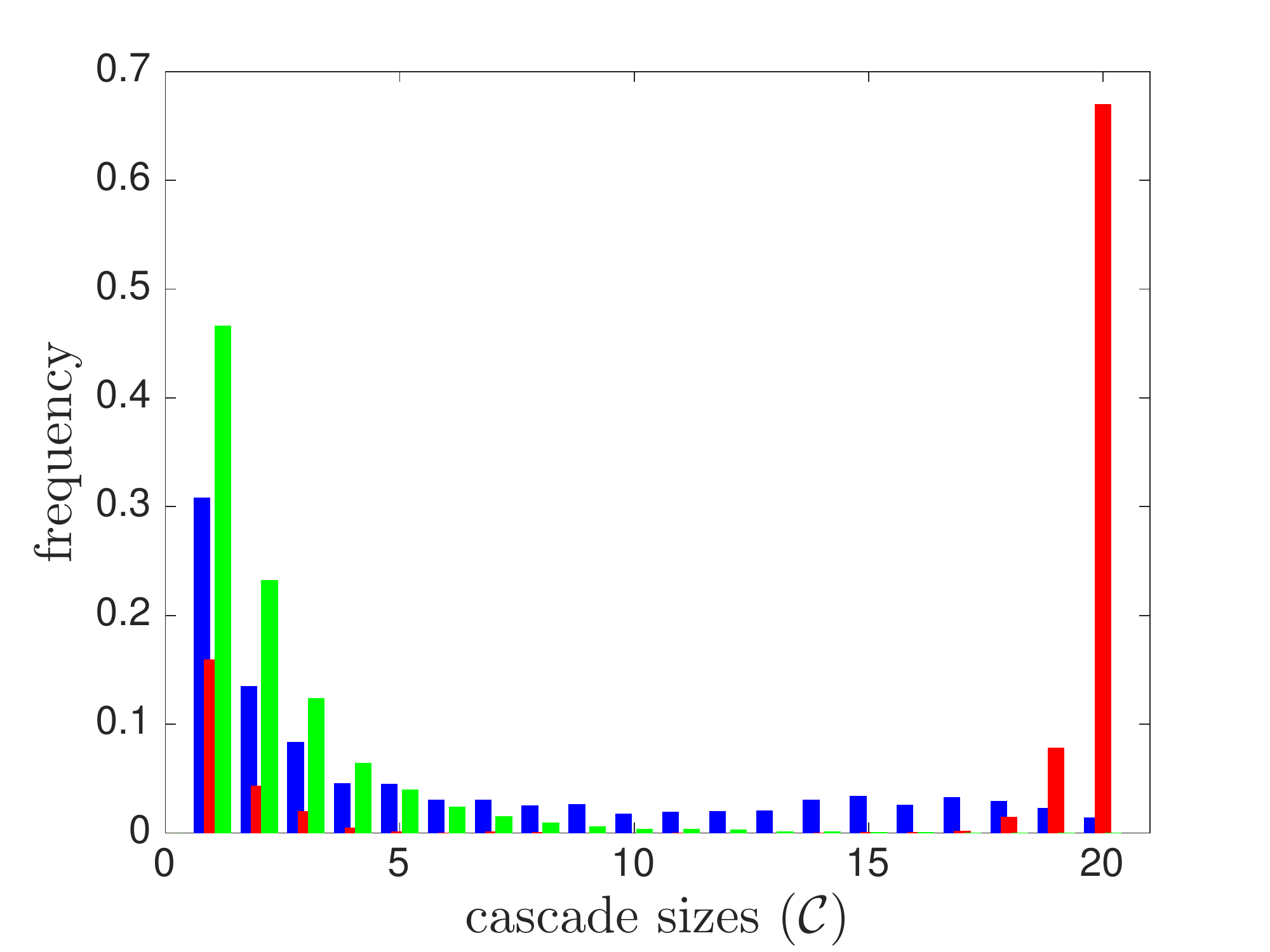}
\text{(b)}
\includegraphics[width=.7\textwidth]{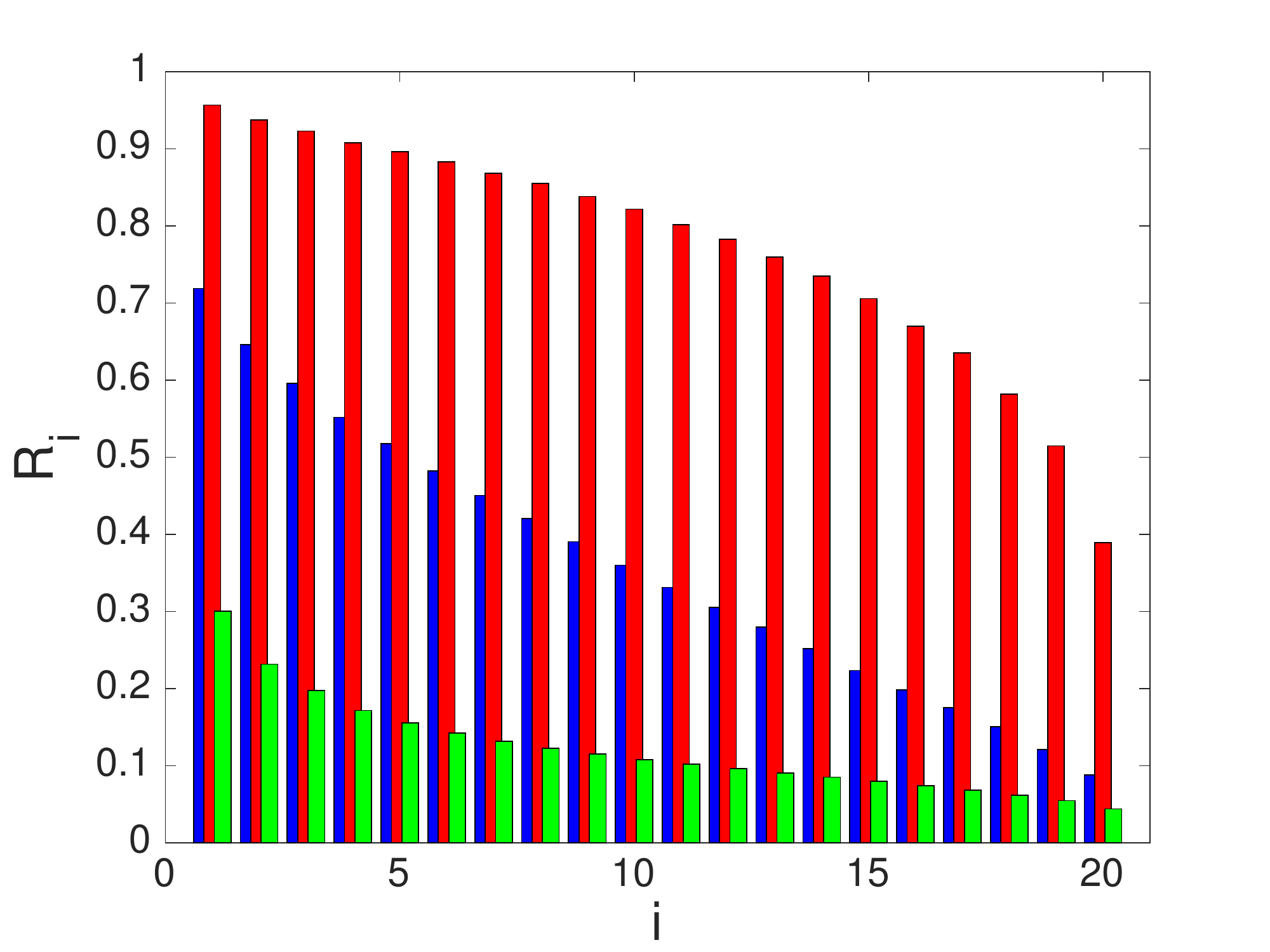}
\text{(c)}
\caption{ {\footnotesize Systemic Risk in $3$ Different Scenarios. \textit{The case of an interbank system composed of $20$ banks without a CDS market is shown in blue. The case of this interbank system with a regulated, covered CDS market is shown in green while the case of this interbank system with an unregulated, naked CDS market is shown in red. Plot (a) shows the histogram of aggregated losses resulting from insolvency cascades in each scenario. Plot (b) shows the histogram of the number of banks that fail in a cascade in each scenario. Plot (c) shows the DebtRanks (i.e. the systemic importance) of the $20$ banks in each scenario.} } }
\label{fig:RiskProfiles}
\end{figure}

In Fig. \ref{fig:RiskProfiles} (a), we plot the aggregate losses to the interbank system resulting from insolvency cascades. These are shown to be considerably decreased when the CDS market is regulated as described previously (shown in green). On the other hand, the situation becomes much worse when the CDS market is unregulated (shown in red) and banks are allowed to speculate by buying `naked' CDSs, i.e. CDSs that are not used as insurance on an underlying loan. We see that the loss distribution is actually bimodal, i.e. there is a strong probability of very high loss, associated with total systemic collapse. Fig. \ref{fig:RiskProfiles} (b) shows the number of banks that fail in a cascade. The patterns are similar to those discussed for (a) above. Fig. \ref{fig:RiskProfiles} (c) shows the corresponding DebtRanks (i.e. the systemic importance) of the $20$ banks in each of the three different scenarios. We can clearly see that by increasing the number of contagion channels, an unregulated CDS market considerably increases the systemic riskiness\footnote{See previous footnote.} of each bank. On the other hand, by properly transferring exposures from one bank to another, the proposed systemic surcharge mechanism for CDSs creates an interbank system in which each institution is considerably less systemically risky.

\begin{figure}[ht] 
  \begin{minipage}[b]
  {0.5\linewidth}
    \includegraphics[width=1.05\linewidth]{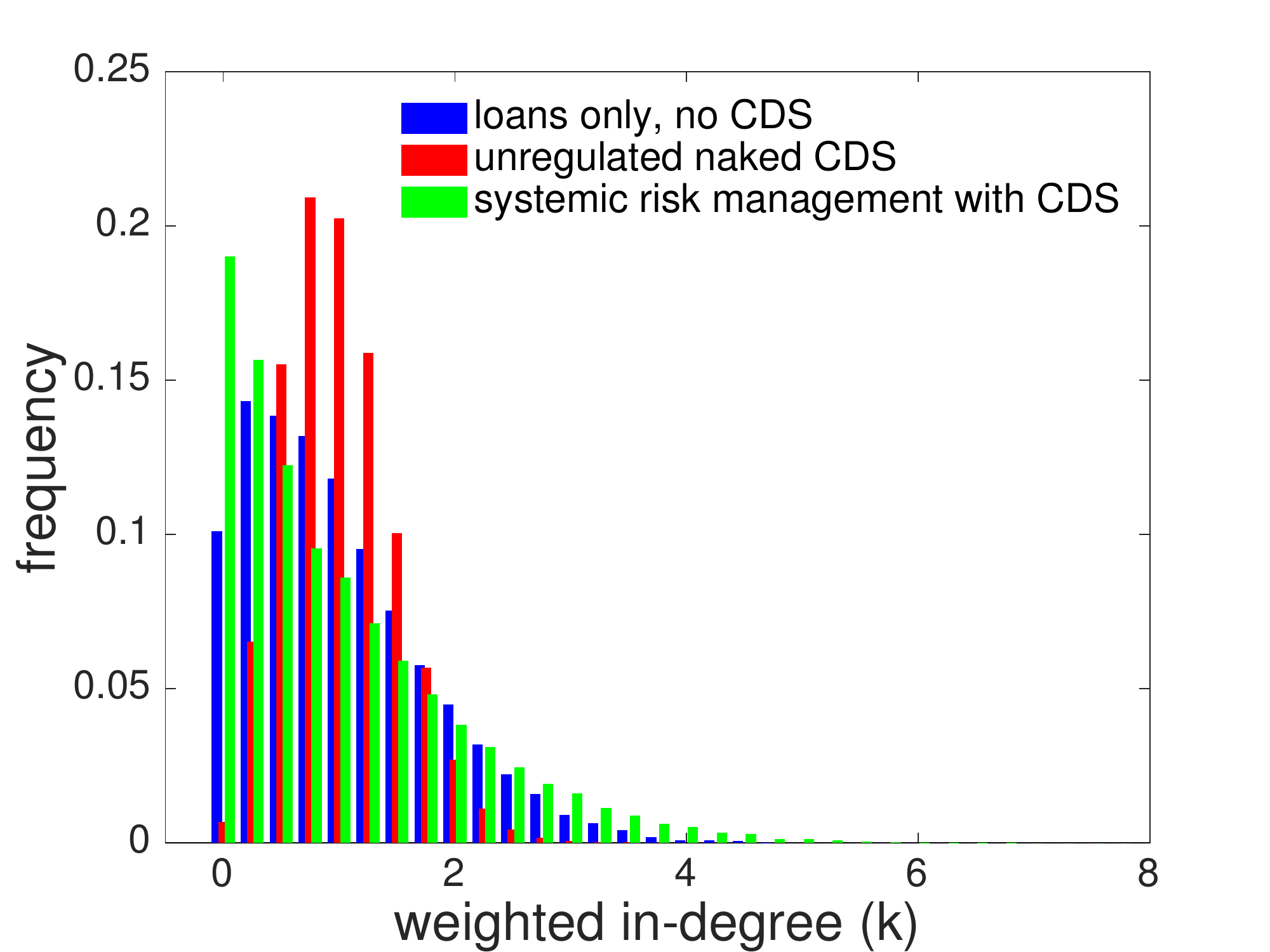}
    \text{(a)}
  \end{minipage} 
   \begin{minipage}[b]{0.5\linewidth}
    \includegraphics[width=1.05\linewidth]{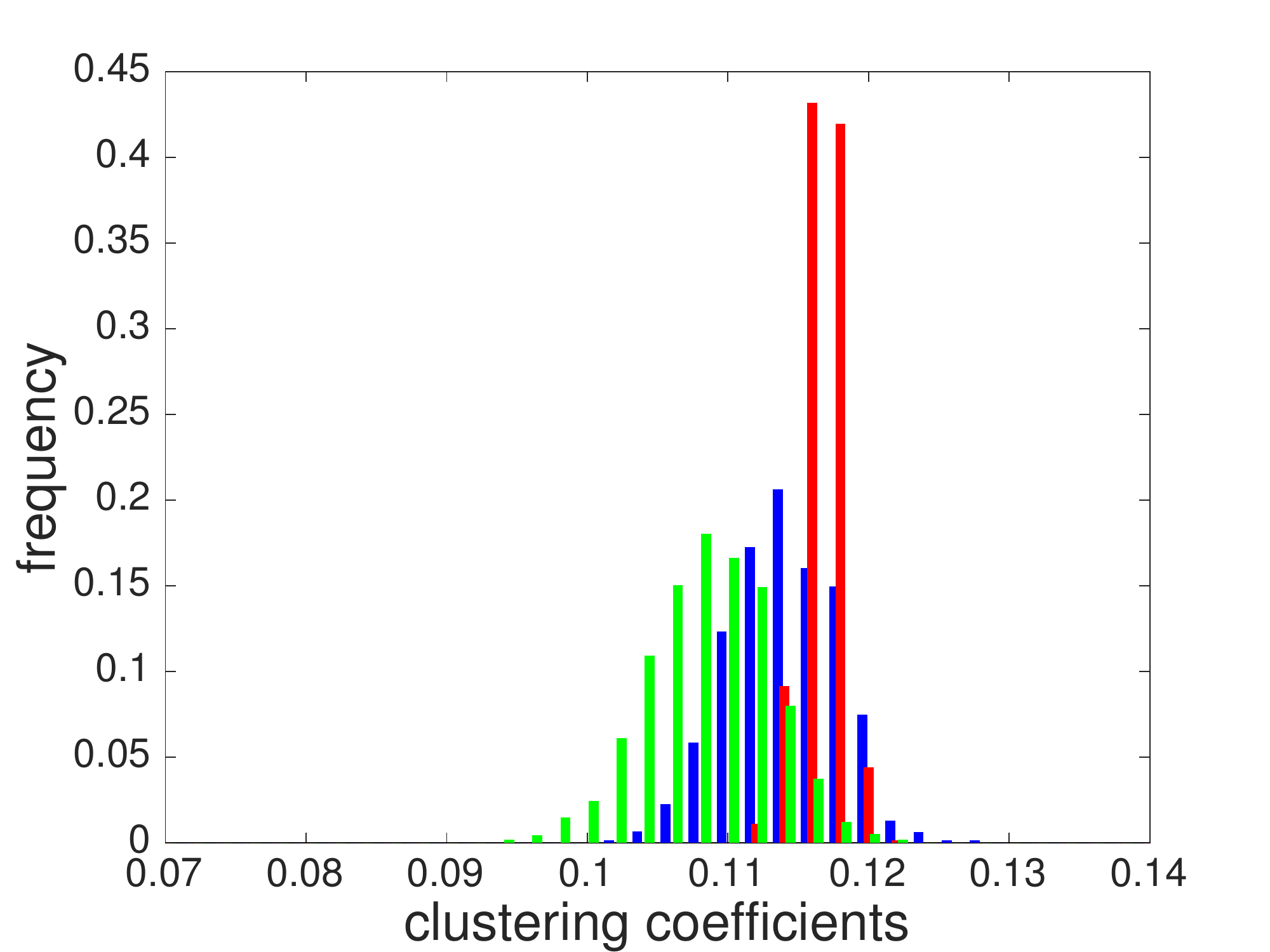} 
    \text{(b)}
  \end{minipage}
  \hfill
    \caption{Network Statistics in $3$ Different Scenarios. \textit{The case of an interbank system composed of $20$ banks without a CDS market is shown in blue. The case of this interbank system with a regulated CDS market is shown in green while the case of this interbank system with an unregulated CDS market is shown in red. Plot (a) shows the empirical weighted in-degree distribution. Plot (b) shows the empirical distribution of the clustering coefficient.}} 
    \label{net_stats} 
\end{figure}

In Fig. \ref{net_stats}, we provide additional network statistics. In Fig. \ref{net_stats}(a), we show the empirical weighted in-degree distribution of the effective exposure network. We can see that a regulated CDS market somewhat shifts the weighted in-degree distribution towards lower degrees. An unregulated CDS market, on the other hand, appears to shift it towards higher degrees and this was to be expected from Fig. \ref{circular_nets}(c), where we saw that many more exposures (edges) were created in the network. The in-degree, however does not give the full picture. Indeed, in the effective exposure network,  it is rather the presence of cycles that can increase the potential size of insolvency cascades by creating long chains of exposures. Since a CDS market regulated with a systemic insurance surcharge was shown to decrease systemic risk, we may expect it to have the effect of cutting cycles of exposures. A statistic that can be used to assess the presence of cycles is the clustering coefficient. We use the {\em weighted clustering coefficient} as defined in \citet{Barrat:2004aa} because the interbank relation network is modeled as a weighted network. In Fig. \ref{net_stats}(b), we see that a regulated CDS market indeed decreases the clustering coefficient. This is achieved in part by encouraging CDS transactions that `cancel' existing exposures, thus decreasing net exposures. An unregulated CDS market, on the other hand, has the effect of increasing the clustering coefficient by creating more exposures and thus more cycles in the effective exposure network.

\section{Conclusion}

In this article, we showed that covered credit default swaps (CDSs) offer a natural way to restructure financial networks, as they can be used to shift loan exposures from one institution to another. When properly regulated, they can lead to network configurations that are more resilient to insolvency cascades. We proposed a taxing mechanism that incentivizes institutions to form CDS contracts that decrease systemic risk by reallocating loan exposures in a more efficient way. A regulator in possession of network information can impose a systemic surcharge to the CDS spread. This surcharge penalizes CDS contracts that increase systemic risk, but leaves other CDS contracts unaffected, thereby effectively rewiring the network of exposures. This systemic surcharge is collected by the regulator and can be used to guarantee the fulfillment of CDS contracts in the event that the seller were to become insolvent. 

We also showed that an unregulated CDS market leads to an interbank system with network properties that are totally different from those of a loans market without derivatives. Unregulated, these credit derivatives can be indeed extremely dangerous as they create additional contagion channels. 

This work stresses the importance of developing regulations that influence the topology of financial  networks. Existing regulations such as Tobin taxes or capital requirements do not affect the structure of financial networks and thus have limited effect on increasing their resilience to insolvency cascades.

\section{Acknowledments}
We thank anonymous referees for the NetSci-X 2016 Conference, January 11-13, 2016, Wroclaw, Poland, where this research was presented, as well as participants for comments.

\section{Declarations of Interest}
We acknowledge financial support from MULTIPLEX, agreement no. 317532 (20).

\bibliographystyle{acmsmall}
\bibliography{acmsmall-sample-bibfile}

\begin{thebibliography}{}

\bibitem[\protect\citeauthoryear{Acemoglu, Ozdaglar, and
  Tahbaz-Salehi}{Acemoglu et~al\mbox{.}}{2015}]{acemoglu2013systemic}
{\sc Acemoglu, D.}, {\sc Ozdaglar, A.}, {\sc and} {\sc Tahbaz-Salehi, A.} 2015.
\newblock Systemic risk and stability in financial networks.
\newblock {\em The american economic review\/}~{\em 105,\/}~2, 564--608.

\bibitem[\protect\citeauthoryear{Amini, Cont, and Minca}{Amini
  et~al\mbox{.}}{2013}]{amini2013resilience}
{\sc Amini, H.}, {\sc Cont, R.}, {\sc and} {\sc Minca, A.} 2013.
\newblock Resilience to contagion in financial networks.
\newblock {\em Mathematical finance\/}.

\bibitem[\protect\citeauthoryear{Aymanns, Georg, and Golub}{Aymanns
  et~al\mbox{.}}{2016}]{aymanns2016illiquidity}
{\sc Aymanns, C.}, {\sc Georg, C.-P.}, {\sc and} {\sc Golub, B.} 2016.
\newblock Illiquidity spirals in over-the-counter repo markets.

\bibitem[\protect\citeauthoryear{Barrat, Barth{\'e}lemy, Pastor-Satorras, and
  Vespignani}{Barrat et~al\mbox{.}}{2004}]{Barrat:2004aa}
{\sc Barrat, A.}, {\sc Barth{\'e}lemy, M.}, {\sc Pastor-Satorras, R.}, {\sc
  and} {\sc Vespignani, A.} 2004.
\newblock The architecture of complex weighted networks.
\newblock {\em Proceedings of the National Academy of Sciences of the United
  States of America\/}~{\em 101,\/}~11, 3747--3752.

\bibitem[\protect\citeauthoryear{Battiston, Puliga, Kaushik, Tasca, and
  Caldarelli}{Battiston et~al\mbox{.}}{2012}]{battiston2012debtrank}
{\sc Battiston, S.}, {\sc Puliga, M.}, {\sc Kaushik, R.}, {\sc Tasca, P.}, {\sc
  and} {\sc Caldarelli, G.} 2012.
\newblock Debtrank: Too central to fail? financial networks, the fed and
  systemic risk.
\newblock {\em Scientific reports\/}~{\em 2}.

\bibitem[\protect\citeauthoryear{Boss, Elsinger, Summer, and Thurner}{Boss
  et~al\mbox{.}}{2004}]{boss2004network}
{\sc Boss, M.}, {\sc Elsinger, H.}, {\sc Summer, M.}, {\sc and} {\sc Thurner,
  S.} 2004.
\newblock Network topology of the interbank market.
\newblock {\em Quantitative Finance\/}~{\em 4,\/}~6, 677--684.

\bibitem[\protect\citeauthoryear{Burkholz, Leduc, Garas, and
  Schweitzer}{Burkholz et~al\mbox{.}}{2016}]{burkholz2015systemic}
{\sc Burkholz, R.}, {\sc Leduc, M.~V.}, {\sc Garas, A.}, {\sc and} {\sc
  Schweitzer, F.} 2016.
\newblock Systemic risk in multiplex networks with asymmetric coupling and
  threshold feedback.
\newblock {\em Physica D: Nonlinear Phenomena\/}~{\em 323}, 64--72.

\bibitem[\protect\citeauthoryear{Cont and Minca}{Cont and
  Minca}{2016}]{cont2014credit}
{\sc Cont, R.} {\sc and} {\sc Minca, A.} 2016.
\newblock Credit default swaps and systemic risk.
\newblock {\em Annals of Operations Research\/}.

\bibitem[\protect\citeauthoryear{Delli~Gatti, Palestrini, Gaffeo, Giulioni, and
  Gallegati}{Delli~Gatti et~al\mbox{.}}{2008}]{delli2008emergent}
{\sc Delli~Gatti, D.}, {\sc Palestrini, A.}, {\sc Gaffeo, E.}, {\sc Giulioni,
  G.}, {\sc and} {\sc Gallegati, M.} 2008.
\newblock {\em Emergent macroeconomics}.
\newblock Springer.

\bibitem[\protect\citeauthoryear{Duffie and Zhu}{Duffie and
  Zhu}{2011}]{duffie2011}
{\sc Duffie, D.} {\sc and} {\sc Zhu, H.} 2011.
\newblock Does a central clearing counterparty reduce counterparty risk?
\newblock {\em Review of Asset Pricing Studies\/}~{\em 1,\/}~1, 74--95.

\bibitem[\protect\citeauthoryear{Eisenberg and Noe}{Eisenberg and
  Noe}{2001}]{eisenberg2001systemic}
{\sc Eisenberg, L.} {\sc and} {\sc Noe, T.~H.} 2001.
\newblock Systemic risk in financial systems.
\newblock {\em Management Science\/}~{\em 47,\/}~2, 236--249.

\bibitem[\protect\citeauthoryear{Elliott, Golub, and Jackson}{Elliott
  et~al\mbox{.}}{2014}]{ElliotGolubJackson2014}
{\sc Elliott, M.}, {\sc Golub, B.}, {\sc and} {\sc Jackson, M.~O.} 2014.
\newblock Financial networks and contagion.
\newblock {\em The American economic review\/}~{\em 104,\/}~10, 3115--3153.

\bibitem[\protect\citeauthoryear{Gaffeo, Gatti, Desiderio, and
  Gallegati}{Gaffeo et~al\mbox{.}}{2008}]{gaffeo2008adaptive}
{\sc Gaffeo, E.}, {\sc Gatti, D.~D.}, {\sc Desiderio, S.}, {\sc and} {\sc
  Gallegati, M.} 2008.
\newblock Adaptive microfoundations for emergent macroeconomics.
\newblock {\em Eastern Economic Journal\/}~{\em 34,\/}~4, 441--463.

\bibitem[\protect\citeauthoryear{Gai and Kapadia}{Gai and
  Kapadia}{2010}]{gai2010contagion}
{\sc Gai, P.} {\sc and} {\sc Kapadia, S.} 2010.
\newblock Contagion in financial networks.
\newblock In {\em Proceedings of the Royal Society of London A: Mathematical,
  Physical and Engineering Sciences}. The Royal Society, rspa20090410.

\bibitem[\protect\citeauthoryear{Gatti, Desiderio, Gaffeo, Cirillo, and
  Gallegati}{Gatti et~al\mbox{.}}{2011}]{gatti2011macroeconomics}
{\sc Gatti, D.~D.}, {\sc Desiderio, S.}, {\sc Gaffeo, E.}, {\sc Cirillo, P.},
  {\sc and} {\sc Gallegati, M.} 2011.
\newblock {\em Macroeconomics from the Bottom-up}. Vol.~1.
\newblock Springer Science \& Business Media.

\bibitem[\protect\citeauthoryear{Gualdi, Tarzia, Zamponi, and Bouchaud}{Gualdi
  et~al\mbox{.}}{2015}]{gualdi2015tipping}
{\sc Gualdi, S.}, {\sc Tarzia, M.}, {\sc Zamponi, F.}, {\sc and} {\sc Bouchaud,
  J.-P.} 2015.
\newblock Tipping points in macroeconomic agent-based models.
\newblock {\em Journal of Economic Dynamics and Control\/}~{\em 50}, 29--61.

\bibitem[\protect\citeauthoryear{Leduc and Thurner}{Leduc and
  Thurner}{2017}]{leduc2015equilibrium}
{\sc Leduc, M.~V.} {\sc and} {\sc Thurner, S.} 2017.
\newblock Incentivizing resilience in financial networks.
\newblock {\em Journal of Economic Dynamics and Control\/}~{\em 82}, 44--66.

\bibitem[\protect\citeauthoryear{Markose}{Markose}{2012}]{markose2012systemic}
{\sc Markose, S.} 2012.
\newblock {\em Systemic risk from global financial derivatives: A network
  analysis of contagion and its mitigation with super-spreader tax}.
\newblock Number 12-282. International Monetary Fund.

\bibitem[\protect\citeauthoryear{Poledna, Molina-Borboa,
  Mart{\'\i}nez-Jaramillo, Van Der~Leij, and Thurner}{Poledna
  et~al\mbox{.}}{2015}]{poledna2015multi}
{\sc Poledna, S.}, {\sc Molina-Borboa, J.~L.}, {\sc Mart{\'\i}nez-Jaramillo,
  S.}, {\sc Van Der~Leij, M.}, {\sc and} {\sc Thurner, S.} 2015.
\newblock The multi-layer network nature of systemic risk and its implications
  for the costs of financial crises.
\newblock {\em Journal of Financial Stability\/}~{\em 20}, 70--81.

\bibitem[\protect\citeauthoryear{Poledna and Thurner}{Poledna and
  Thurner}{2016}]{poledna2014elimination}
{\sc Poledna, S.} {\sc and} {\sc Thurner, S.} 2016.
\newblock Elimination of systemic risk in financial networks by means of a
  systemic risk transaction tax.
\newblock {\em Quantitative Finance\/}~{\em 16,\/}~10, 1599--1613.

\bibitem[\protect\citeauthoryear{Schuldenzucker, Seuken, and
  Battiston}{Schuldenzucker et~al\mbox{.}}{2016}]{schuldenzucker2016clearing}
{\sc Schuldenzucker, S.}, {\sc Seuken, S.}, {\sc and} {\sc Battiston, S.} 2016.
\newblock Clearing payments in financial networks with credit default swaps
  [extended abstract].
\newblock In {\em Proceedings of the 17th ACM Conference on Economics and
  Computation}. EC '16. ACM, Maastricht, The Netherlands, 759--759.

\bibitem[\protect\citeauthoryear{Thurner and Poledna}{Thurner and
  Poledna}{2013}]{thurner2013debtrank}
{\sc Thurner, S.} {\sc and} {\sc Poledna, S.} 2013.
\newblock Debtrank-transparency: Controlling systemic risk in financial
  networks.
\newblock {\em Scientific reports\/}~{\em 3}.

\end{thebibliography}

\end{document}